\documentclass[doublecol]{epl2} 

\usepackage{graphicx}
\usepackage{subcaption}
\usepackage[american]{babel}
\usepackage{hyperref}	          
\usepackage{epstopdf}

\usepackage[nodisplayskipstretch]{setspace}

\begin{document}

\title{A lattice Boltzmann method based on generalized polynomials and its application for electrons in metals}

\author{Rodrigo C. V. Coelho\inst{1}\thanks{\email{rcvcoelho@if.ufrj.br}}, Anderson S. Ilha\inst{2}\thanks{\email{anderson.ilha@gmail.com}}, Mauro M. Doria\inst{1,3}\thanks{\email{mmd@if.ufrj.br}} }
\shortauthor{R. C. V. Coelho \etal}

\institute{
\inst{1} Departamento de F\'{\i}sica dos S\'{o}lidos, Universidade Federal do Rio de Janeiro, 21941-972 Rio de Janeiro, Brazil\\
\inst{2} Instituto Nacional de Metrologia, Normaliza\c{c}\~ao e Qualidade Industrial, Duque de Caxias 25.250-020 RJ Brazil\\
\inst{3} Dipartimento di Fisica, Universit\`a di Camerino, I-62032 Camerino, Italy}


\date{\today}

\abstract{A lattice Boltzmann method is proposed based on the expansion of the equilibrium distribution function in powers of a new set of generalized orthonormal polynomials which are here presented. The new polynomials are orthonormal under the weight defined by the equilibrium distribution function itself. The D-dimensional  Hermite polynomials is a sub-case of the present ones, associated to the particular weight of a gaussian function. The proposed lattice Boltzmann method allows for the treatment of semi-classical fluids, such as electrons in metals under the Drude-Sommerfeld model, which is a particular case that we develop and validate by the Riemann problem.}


\maketitle
%
\textbf{Introduction}. -- The {\it lattice Boltzmann method} (LBM) has been applied in many problems of classical fluid dynamics ranging from biology to material science~\cite{succi08}. Its strength lies in the ability to easily incorporate complex physical phenomena, naturally present in micron scale, porous media, and multiphase flows. The method is implemented on a regular discrete position space, such that the continuum fluid mechanics is retrieved at distances much larger than the lattice scale. Initially conceived as a gas of colliding particles with confined motion along a  lattice, the LBM quickly reached widespread use once its underlying mathematical structure was understood. The LBM was found to rely on the {\it D-dimensional Hermite polynomials} (DHP) and the {\it Maxwell-Boltzmann} (MB) {\it equilibrium distribution function} (EDF)~\cite{shan98,luo98}. It was H. Grad~\cite{grad49,kremer10} who first  used the DHP to describe the microscopic velocity space~\cite{kremer10} of the Boltzmann equation, but these polynomials are important in other areas as well, such as quantum optics~\cite{kok01}. The LBM has been mostly applied to the MB isothermal fluid~\cite{ansumali05,chikatamarla06}, but recent applications to explain relativistic flows~\cite{romatschke11} and graphene~\cite{mendoza13} reveal a new trend in the LBM, which is to go beyond the standard framework of the DHP and the MB distribution.

In this letter we propose a novel LBM for the treatment of semiclassical fluids that incorporates the previous LBM as a sub-case, and because of its convergent properties can go beyond in case of the above fluids, specially in situations under complexity.  The present advancements are due to our discovery of a new class of D-dimensional polynomials.
Hence the present proposal opens the way to treat electrons in metals, for instance, through the LBM method, a long expected goal not yet achieved for lackness of the appropriate theoretical tools which are here presented. It is well-known that the Boltzmann-BGK equation provides the framework to understand the Drude-Sommerfeld model which is the standard textbook procedure to treat electrons in metals ~\cite{ashcroft76}. Electrons form a gas, similarly to the atoms in a rarefied gas, however at room temperature they are essentially governed by their zero temperature properties, and so the Fermi velocity $v_F$, and not the sound velocity, $c_r$, is important (for instance, for cooper~\cite{ashcroft76}, $v_F = 1.57 \times 10^{6} $ m/s). The displacement velocity is very low in metals, for typical electric fields, $u \sim 0.1\mbox{-}1.0$ m/s, that is, $u/v_F \sim 6.4\mbox{-}64\times  10^{-8}$, while  Mach's number diverges for $T=0$. The new polynomials bring a new  velocity scale, $c_0$, that can even be {\it local} whereas $c_r$ is necessarily global, and so, provide clear advantage over previous methods. The novel LBM will bring advances in areas such as microscale electronics since it allows for the treatment of the electron gas in any geometrical arrangement under the presence of inhomogeneous fields, temperature gradients and defects. Since the electronic mean free path $l$ falls in the sub-micrometer range (the collision time of cooper~\cite{ashcroft76} is $\tau = 2.1 \times 10^{-13}$ s and $l=v_F\tau \sim  0.33$ $\mu$m) the macroscopic equations of motion of continuum fluid mechanics obtained from this novel LBM become useful~\cite{coelho14} (small Knudsen number, $Kn \sim l/L$, $L$ is the characteristic length scale of the system). 

Semi-classical fluids have particles that obey either the Fermi-Dirac (FD) or the Bose-Einstein (BE) statistics, and only for low densities and high temperatures become the well-known MB statistics satisfied by the atoms in a gas. Long ago E. A. Uheling and G. E. Uhlenbeck~\cite{uehling33} generalized the Boltzmann equation to account for particles obeying the BE-FD statistics. For all three statistics the scattering by collisions is sufficiently well described by the Bhatnagar-Gross-Krook (BGK) term~\cite{bgk54,lutsko97} and the LBM is essentially a numerical implementation of the Bolztmann-BGK equation~\cite{bosch13}. Quantum~\cite{yang10,coelho14} and classical particles under thermal gradients~\cite{philippi06}, have been treated by the LBM in the context of the DHP, which renders to them a very restrictive window of convergence since their EDFs are expanded around the gaussian function. We show here that significant improvement in convergence of the LBM, and therefore usefulness in its practical descriptions, is obtained if the EDF is expanded near to itself, for a shifted argument, which leads to the new set of D-dimensional polynomials proposed in this letter. The new polynomials are orthonormal under a weight given by the EDF itself, whereas the DHP is fixed to the gaussian weight. The isothermal MB LBM is convergent for a small Mach number ($Ma \sim u/c_r$, $u$ is the displacement velocity and $c_r\equiv\sqrt{kT_r/m}$ is essentially the speed of sound at temperature $T_r$), whereas the the present LBM is convergent for small $u/c_0$. 

The EDF depends explicitly on the microscopic velocity, $\boldsymbol{\xi}$, and implicitly on the position, $\boldsymbol{x}$, through the local macroscopic parameters, such as the density $\rho(\boldsymbol{x})$ (or the chemical potential $\mu(\boldsymbol{x})$), the macroscopic (displacement) velocity $\boldsymbol{u}(\boldsymbol{x})$, and the temperature $\theta(\boldsymbol{x})$. The FD ($+$) and the BE ($-$) EDFs  correspond to $f^{(eq)}_{\,FD/BE} = \left[ \exp{\left(-\mu/\theta\right)}\exp\left(\boldsymbol{\xi}^2/2 \theta\right)\pm1\right]^{-1}$, and the MB is $f^{(eq)}_{MB} = \rho_0 \exp\left( -\boldsymbol{\xi}^2/2 \theta\right )/(2 \pi \theta )^{D/2}$, where $\rho_0$ is a dimensionless density. All variables are defined dimensionless $(m=k_B=c=\hbar=1)$ by means of the scale set by $T_r$, which fixes $c_r$. The Chapman-Enskog analysis applies~\cite{coelho14}, which means that the first three moments are either computable from the non-equilibrium distribution function or from the known EDF~\cite{wagner08}. The first three moments are given by, $\rho \equiv \int d^{D}\boldsymbol{\xi}\, f^{(eq)}(\boldsymbol{\xi}- \boldsymbol{u})$, $\rho \boldsymbol{u} = \int d^{D}\boldsymbol{\xi}\, \boldsymbol{\xi} \, f^{(eq)}(\boldsymbol{\xi}- \boldsymbol{u})$ and $D\rho \bar \theta \equiv \int d^{D}\boldsymbol{\xi}\, \left( \boldsymbol{\xi}- \boldsymbol{u} \right)^2  \, f^{(eq)}(\boldsymbol{\xi}-\boldsymbol{u})$, where the third moment is the internal energy, defined through the pseudo-temperature~\cite{coelho14}, which in the classical case becomes the temperature, $\bar \theta = \theta$.

\textbf{Generalized polynomials in D dimensions}. --  Consider the D-dimensional space of the microscopic velocity, $\boldsymbol{\xi} \equiv (\xi_1,\xi_2,\cdots,\xi_D)$, endowed with a weight function that can be, for instance, any one of the three EDF's previously discussed: $\omega=\left (f^{(eq)}_{FD},\,f^{(eq)}_{BE},\,f^{(eq)}_{MB}\right )$. We claim here the existence of a set of orthonormal polynomials $\mathcal{P}_{i_1\cdots i_N}( \boldsymbol{\xi})$, such that,
\begin{equation}\label{omeg-feq}
\int d^D \boldsymbol{\xi} \, \omega( \boldsymbol{\xi} )\mathcal{P}_{i_1\cdots i_N}( \boldsymbol{\xi})\mathcal{P}_{j_1\cdots j_M}(\boldsymbol{\xi})=
\delta_{\scriptscriptstyle {N M}}\delta_{i_1\cdots i_N\vert j_1\cdots j_M},
\end{equation}
where $\omega(\boldsymbol{\xi} ) \equiv f^{(eq)}(\boldsymbol{\xi}, \boldsymbol{u} = 0)$. The DHP correspond to the gaussian function $\omega(\boldsymbol{\xi}) =  \exp(-\boldsymbol{\xi}^2/2)/(2 \pi )^{D/2}$. In fact the existence of the generalised polynomials only relies on some general properties of the weight function, such as dependence on $\xi \equiv\vert \boldsymbol{ \xi} \vert$ and its  evanishment at extremely high microscopic velocities, $\omega(\boldsymbol{\xi}) \rightarrow 0$ for $ \xi \rightarrow \infty$, such that the integrals $I_{N}$ given below, are well defined.
\begin{eqnarray}\label{i2n-cont}
\int d^D \boldsymbol{\xi} \, \omega( \boldsymbol{\xi} ) \, \xi_{i_1}\cdots \xi_{i_{N}}=  I_{N}\,\delta_{i_1\cdots i_{N}},
\end{eqnarray}
By symmetry it holds that $I_N=0$ for $N$ odd. We introduce the following tensors which are sums of products of the Kronecker's delta function ($\delta^{ij}=1$ for $i=j$ and 0 for $i\neq j$), namely, $\delta_{i_1\cdots i_N\vert j_1\cdots j_N} \equiv \delta_{i_1 j_1}\cdots \delta_{i_Nj_N}\,+$ all permutations  of  $j$'s and $\delta_{i_1\cdots i_N\, j_1\cdots j_N} \equiv \delta_{i_1 j_1}\cdots \delta_{i_Nj_N}+$ all permutations. Thus they contain $N\,\!!$ and $(2N)\,\!!$ terms, respectively. Using the aforementioned symmetries the $I_{2N}$ are given by,
\begin{eqnarray}\label{i2n-cont2}
I_{2N}=\frac{\pi^{\frac{D}{2}}}{2^{N-1}\Gamma\big(N+\frac{D}{2}\big)}\int_{0}^{\infty}d\xi\, \omega(\xi)\,\xi^{2N+D-1}.
\end{eqnarray}
A direct consequence of eq.(\ref{omeg-feq}) and (\ref{i2n-cont}) is that the moments are given by $\rho=I_0$ and $\bar \theta=I_2/I_0$.
Like the DHP the new polynomials $\mathcal{P}_{i_1 \cdots i_N}(\boldsymbol{\xi})$ are symmetrical in the indices $i_1 \cdots i_N$, have the parity property $\mathcal{P}_{i_1 \cdots i_N}(-\xi_{i_1},\ldots, -\xi_{i_k},\ldots,-\xi_{i_N})= (-1)^{N}\mathcal{P}_{i_1 \cdots i_N}(\xi_{i_1},\ldots, \xi_{i_k},\ldots \xi_{i_N})$ and are tensors in Euclidean space expressed in terms of the vector components $\xi_{i}$ and of $\delta_{ij}$.
However they have a new property not present in the DHP, namely, they are of N$^{\mbox{th}}$ order in $\xi$. This is an important feature that shows that DHP are a sub-class of the new polynomials and not vice-versa.
The first four ones are given by,
\begin{eqnarray}
\mathcal{P}_0(\boldsymbol{\xi}) = c_0,\label{H0}
\end{eqnarray}
\begin{eqnarray}
\mathcal{P}_{i_1}(\boldsymbol{\xi})=c_1\,\xi_{i_1},\label{H1}
\end{eqnarray}
\begin{eqnarray}
\mathcal{P}_{i_1 i_2}(\boldsymbol{\xi})=c_2\,\xi_{i_1} \xi_{i_2} + \left( {\bar c}_2 \xi^2+ {c^\prime}_2 \right)\,\delta_{i_1 i_2}, \label{H2}
\end{eqnarray}
\begin{eqnarray}
&&\mathcal{P}_{i_1 i_2 i_3}(\boldsymbol{\xi})=c_3 \,\xi_{i_1} \xi_{i_2}\xi_{i_3}   +\left( {\bar c}_3 \xi^2+ {c^\prime}_3 \right)\, \big (\xi_{i_1} \delta_{i_2 i_3}\nonumber \\
&& + \xi_{i_2}\delta_{i_1 i_3} +\xi_{i_3} \delta_{i_1 i_2}\big), \label{H3} \: \mbox{and},
\end{eqnarray}
\begin{eqnarray}
&&\mathcal{P}_{i_1 i_2 i_3 i_4}(\boldsymbol{\xi})=c_4 \,\xi_{i_1} \xi_{i_2}\xi_{i_3} \xi_{i_4}+  \left( {\bar c}_4 \xi^2+ {c^\prime}_4 \right)\,\left (\xi_{i_1}\xi_{i_2}
\delta_{i_3 i_4}  \right . \nonumber \\
&& \left . + \xi_{i_1}\xi_{i_3}
\delta_{i_2 i_4}+\xi_{i_1}\xi_{i_4}
\delta_{i_2 i_3}+\xi_{i_2}\xi_{i_3} \delta_{i_1 i_4}+ \xi_{i_2}\xi_{i_4}
\delta_{i_1 i_3}\right .\nonumber \\
&&\left .+\xi_{i_3}\xi_{i_4}\delta_{i_1 i_2} \right )+\left( {\bar d}_4 \xi^4+{d^\prime}_4 \xi^2+  {d}_4 \right)\,\delta_{i_1 i_2 i_3 i_4}.\label{H4}
\end{eqnarray}
All the coefficients are solely functions of the integrals $I_{2N}$:
$c_K=1/\sqrt{I_{2K}}$ for $K=0,1,2,3,$ and $4$, and ${c^\prime}_K=-c_K\big(I_{2K-2}/I_{2K-4}\big)\Delta_{2K-2}$, ${\bar c}_K=c_K\big(-1+\Delta_{2K-2} \big)/\big(D+2K-4 \big)$ for $K=1,2,3,$ and $4$. We define for $L$ even, $\Delta_L^2\equiv 2/\left[\big(D+L\big)-J_L\big(D+L-2\big)\right]$ and $J_{L}\equiv I_L^2/I_{L+2}I_{L-2}$.
The remaining coefficients are 
\begin{eqnarray}
d_4^2=\frac{8\delta_4^2I_4}{\delta_2\left[\delta_2\delta_6\left(D+4\right)-\delta_4^2 D \right]},
\end{eqnarray}
\begin{eqnarray}
d^\prime_4= -\frac{d_4}{D}\left[\frac{I_0}{I_2}+\frac{I_4\delta_2}{I_2\delta_4}\right]+\frac{2c_4I_6\Delta_6}{D I_4}, \:\:\mbox{and}
\end{eqnarray}
\begin{eqnarray}
{\bar d}_4 = \frac{d_4 \delta_2}{ D\left(D+2\right)\delta_4}+\frac{c_4\left[D-2\left(D+2\right) \Delta_6 \right]}{D\left(D+2\right)\left(D+4\right)},
\end{eqnarray}
where $\delta_L \equiv 2I_{L+2}I_{L-2}/\Delta_L^2$.

For the FD (+) and BE (-) weights one obtains that
$I_{2N}=(2\pi)^{D/2}\theta^{N+D/2}g_{(N+D/2)}\left(e^{\mu/\theta}\right)$, where $g_{\nu} (z) =\int_0^{\infty}dx\, x^{\nu-1}/(z^{-1}e^x\pm1)$ is the FD (+) or BE (-) integral. The FD weight in the so-called Sommerfeld's limit~\cite{pathria11}, used for the treatment of electrons in metals, becomes,
\begin{eqnarray}\label{sommerfeld}
&&I_{2N}=\frac{(2\pi)^{\frac{D}{2}}\mu^{N+\frac{D}{2}}}{\Gamma\left( N+ \frac{D}{2}+1\right)} \\
&&\cdot\left[1+\frac{\pi^2}{6}\left(N+\frac{D}{2}\right)\left(N+\frac{D}{2}-1 \right)\left(\frac{\theta}{\mu} \right)^2+\cdots\right]\nonumber,
\end{eqnarray}
where $\theta/\mu <<1 $ such that terms of order $(\theta/\mu)^4$ and higher are disregarded.
For the MB weight $I_{2N}=I_0\theta^N$ and $\mathcal{P}_{i_1 i_2 \cdots i_N}(\boldsymbol{\xi})=\mathcal{H}_{i_1 i_2 \cdots i_N}(\boldsymbol{\xi}/\sqrt{\theta})/\sqrt{I_0}$, where $\mathcal{H}_{i_1 i_2 \cdots i_N}$ are the DHP. In the Gaussian limit $I_{2N}=1$, and so the coefficients proportional to $\xi^2$ and $\xi^4$, that multiply the tensors $(\xi_{i_1}\xi_{i_1}\cdots)\delta_{i_3i_4\cdots}$, vanish. The remaining ones becomes equal to $\pm 1$: $c_K=1$, $\vert { c^\prime}_K \vert =1$, ${\bar c}_K = 0$, $\vert {d_4} \vert = 1$, ${\bar d}_4= 0$, and $ {d^\prime}_4=0$.\\

\textbf{The EDF expansion in generalized polynomials}. --  Consider an EDF expanded in the standard way, where the coefficients are determined by the orthonormality of the polynomial basis,
\begin{eqnarray}\label{expansao0}
f^{(eq)}(\boldsymbol{\xi} -\boldsymbol{u})= \omega( \boldsymbol{\xi} ) \sum_{N=0}^{K}\frac{1}{N!} \mathcal{A}_{i_1\,i_2 \cdots i_N}(\boldsymbol{u} ) \mathcal{P}_{i_1\,i_2\cdots i_N}( \boldsymbol{\xi}),
\end{eqnarray}
where
\begin{eqnarray}
\mathcal{A}_{i_1\,i_2 \cdots i_N}(\boldsymbol{u} )= \int d^D \boldsymbol{\xi} \, f^{(eq)}(\boldsymbol{\xi}-\boldsymbol{u}) \mathcal{P}_{i_1\,i_2\cdots i_N}( \boldsymbol{\xi}).\nonumber
\end{eqnarray}
The advantages of the new polynomials over the DHP are clearly seen in the above expression:
$f^{(eq)}\left({\boldsymbol{\xi}}-{\boldsymbol{u}}\right)  = \omega\left({\boldsymbol{\xi}}\right)\cdot S({\boldsymbol{\xi}},{\boldsymbol{u}})$.
In case $\omega \equiv f^{(eq)}$, the expansion $S({\boldsymbol{\xi}},{\boldsymbol{u}}\rightarrow 0)\rightarrow 1$, which shows that corrections in ${\boldsymbol{u}}$ are necessarily small. Nevertheless this is not the case for the DHP, where $\omega$ is the gaussian and $f^{(eq)}$ is the FD-BE EDF, for instance.
Thus assume that the EDF is the weight itself and define  $\boldsymbol{\eta}\equiv\boldsymbol{\xi}-\boldsymbol{u}$ to obtain that,
\begin{eqnarray}\label{expansao1}
\mathcal{A}_{i_1\,i_2 \cdots i_N}(\boldsymbol{u} )= \int d^D \boldsymbol{\eta} \, \omega(\boldsymbol{\eta}) \mathcal{P}_{i_1\,i_2\cdots i_N}( \boldsymbol{\eta}+\boldsymbol{u}),
\end{eqnarray}
It is advantageous to expand the polynomial $\mathcal{P}_{i_1\,i_2\cdots i_N}( \boldsymbol{\eta}+\boldsymbol{u})$ in a sum over polynomials $\mathcal{P}_{i_1\,i_2\cdots i_M}( \boldsymbol{\eta})$ of equal or lower order ($M \le N$). This is because, according to Eq.(\ref{omeg-feq}), only the term proportional to $\mathcal{P}_0(\boldsymbol{\eta})$ matters, since only the lowest order term is a constant. Therefore from  $\mathcal{P}_{i_1\,i_2\cdots i_N}( \boldsymbol{\eta}+\boldsymbol{u})= \cdots + \mathcal{P}^\prime_{i_1\,i_2\cdots i_N}(\boldsymbol{u})\mathcal{P}_0(\boldsymbol{\eta})/c_0$, where $\mathcal{P}^\prime$ is a polynomial of the same order of $\mathcal{P}$ but with distinct coefficients, it is straightforward to obtain that
$\mathcal{A}_{i_1\,i_2 \cdots i_N}(\boldsymbol{u})=\mathcal{P}^\prime_{i_1\,i_2 \cdots i_N}(\boldsymbol{u})I_0$. The first four coefficients directly obtained from Eq.(\ref{expansao1}) are given by,
\begin{eqnarray}
\mathcal{A}_0(\boldsymbol{u}) = I_0 c_0, \label{A0}
\end{eqnarray}
\begin{eqnarray}
\mathcal{A}_{i_1}(\boldsymbol{u})=I_0 c_1\,u_{i_1},\label{A1}
\end{eqnarray}
\begin{eqnarray}
\mathcal{A}_{i_1 i_2}(\boldsymbol{u})= I_0 \left ( c_2 u_{i_1} u_{i_2} + {\bar c}_2 \boldsymbol{u}^2 \,\delta_{i_1 i_2} \right ),\label{A2}
\end{eqnarray}
\begin{eqnarray}
&&\mathcal{A}_{i_1 i_2 i_3}(\boldsymbol{u})=I_0\{ c_3 \,u_{i_1} u_{i_2}u_{i_3}  +\big [ {c^\prime}_3 \left(1-J_2 \right) \nonumber \\
&&+{\bar c}_3\boldsymbol{u}^2\big]\left (u_{i_1} \delta_{i_2 i_3} + u_{i_2}\delta_{i_1 i_3} +u_{i_3} \delta_{i_1 i_2}\right ) \}, \: \mbox{and},  \label{A3}
\end{eqnarray}
\begin{eqnarray}
&&\mathcal{A}_{i_1 i_2 i_3 i_4}(\boldsymbol{u})=I_0\Big\{{c}_4 \,u_{i_1} u_{i_2}u_{i_3} u_{i_4}+ \big [\left( 1-J_2J_4\right){c^\prime}_4\nonumber \\
&&+{\bar c}_4 \boldsymbol{u}^2\big ](u_{i_1}u_{i_2}\delta_{i_3 i_4} + u_{i_1}u_{i_3}
\delta_{i_2 i_4}+u_{i_1}u_{i_4}\delta_{i_2 i_3} \nonumber \\
&& +u_{i_2}u_{i_3} \delta_{i_1 i_4} +u_{i_2}u_{i_4} \delta_{i_2 i_3}+u_{i_3}u_{i_4}\delta_{i_1 i_2}) \\ \label{A4}
&&+\left [ \left(2\frac{I_2}{I_0}\left( \bar c_4+(D+2)\bar d_4\right) +d^\prime_4\right ) \boldsymbol{u}^2+\bar d_4 \boldsymbol{u}^4  \right ]\,\delta_{i_1 i_2 i_3 i_4}\Big \}.\nonumber  
\end{eqnarray}
It can be proven that $\mathcal{A}_{i_1i_2\cdots i_{N}}(\boldsymbol{u}=0)=0$ for $N>0$ by taking  $\boldsymbol{u}=0$ in eq.(\ref{expansao0}).
\begin{eqnarray}\label{expansao2}
\sum_{N=0}^{K}\frac{1}{N!} \mathcal{A}_{i_1\,i_2 \cdots i_N}(\boldsymbol{u}=0 ) \mathcal{P}_{i_1\,i_2\cdots i_N}( \boldsymbol{\xi})=1.
\end{eqnarray}
Since $\mathcal{A}_0\mathcal{P}_0=1$ from eqs.(\ref{H0}) and (\ref{A0}), all the remaining coefficients must vanish.

The decomposition of the EDF in terms of the new set of polynomials displays notorious advantages over any another choice of weight. The displacement velocity controls the smallness of the coefficients $\mathcal{A}_{i_1\,i_2 \cdots i_N}$, which to the lowest order are linear and quadratic in $\boldsymbol{u}$ for the odd and even ($N>0$) coefficients, respectively.
To achieve conservation of mass and momentum it is enough to carry the expansion of eq.(\ref{expansao0}) to order $N=3$, and to include energy conservation one order further, $N=4$, and no more~\cite{coelho14}. Thus we find that convergence is achieved in such orders, just demanding a small ratio $u/c_0$. Recall that a distinct choice of weight can render coefficients that do not necessarily vanish for $\boldsymbol{u}=0$, and so, have other parameters that control their smallness. This is the case of the thermal LBM developed over the gaussian weight, whose coefficients also depends on the temperature deviation~\cite{philippi06,coelho14} from a reference value, $\theta-1$.

The generalised polynomials makes the EDF a function of powers of the ratio $u/c_0$ instead of $u/c_r$, but notice that up to this point both $\boldsymbol{u}$ and $\boldsymbol{\xi}$ are normalized by $c_r$. A glimpse into this change of scale is provided by looking at  the highest order term in the velocities of the polynomials, $\mathcal{P}_{i_1\,i_2\cdots i_N}( \boldsymbol{\xi})= (1/\sqrt{I_{2N}})\xi_{i_1} \xi_{i_2}\cdots \xi_{i_N}+\ldots$, and the coefficients,
$\mathcal{A}_{i_1\,i_2\cdots i_N}(\boldsymbol{u})= (I_0/\sqrt{I_{2N}}) u_{i_1} u_{i_2}\cdots u_{i_N}+\ldots$. The presence of the $I_{2N}$ introduces the scale $c_0= (I_{2N})^{1/2N}$, and we define $\boldsymbol{\xi^\prime}\equiv \boldsymbol{\xi}/c_0$  and $\boldsymbol{u^\prime}\equiv \boldsymbol{u}/c_0$ such that $\mathcal{P}_{i_1\,i_2\cdots i_N}( \boldsymbol{\xi})=\xi^\prime_{i_1} \xi^\prime_{i_2}\cdots \xi^\prime_{i_N}+\ldots$,and,
$\mathcal{A}_{i_1\,i_2\cdots i_N}(\boldsymbol{u})=I_0 u^\prime_{i_1} u^\prime_{i_2}\cdots u^\prime_{i_N}+\ldots$.
Consequently $c_r$ is no longer present, as can be checked,
for instance, in the Sommerfeld limit of the FD weight, Eq.(\ref{sommerfeld}), since $c_0\sim \sqrt{\mu}$ and because $\mu$ is also normalized by $c_r^2$.  The overall result is that the chemical potential sets the new velocity scale.\\

\textbf{The Gaussian quadrature}. --  The Gaussian quadrature provides a way to calculate the integral of a function as a sum over some of its values at a set of points $\alpha$ multiplied by pre-determined weights $w_{\alpha}$. We claim here that the Gaussian quadrature holds for $\omega(\boldsymbol{\xi} ) = f^{(eq)}(\boldsymbol{\xi}, \boldsymbol{u}=0)$, such that $\int d^D \boldsymbol{\xi}\,\omega(\boldsymbol{\xi})\, g(\boldsymbol{\xi})= \sum_{\alpha}w_{\alpha}g_{\alpha}$, $g_{\alpha} \equiv g(\boldsymbol{\xi}_{\alpha})$. Thus Eqs.(\ref{i2n-cont2}) turn into the following relations.
\begin{eqnarray}\label{i2n-disc}
\sum_{\alpha} w_{\alpha} \,\xi_{\alpha i_1} \cdots \xi_{\alpha i_{N}}  = I_{N}\, \delta_{i_1\cdots i_{N}},
\end{eqnarray}
such that $I_N=0$ for $N$ odd. The passage from the continuous to the discrete consists in determining lattices, namely the sets of $\xi_{\alpha i_1}$ and $w_{\alpha}$ that satisfy Eqs.(\ref{i2n-disc}) once known the integrals $I_{2N}$ of Eqs.(\ref{i2n-cont}). In the present approach the weights becomes function of the above integrals, $w_{\alpha}(I_{2N})$, and must be updated at each time step of the evolution procedure.

The normalization of the velocities changes from $c_r$ to $c_0$ once the lattice of {\it geometrical velocities}, defined by the set of  $\alpha$ vectors  $\boldsymbol{e}_{\alpha}\equiv\boldsymbol{\xi}_{\alpha}/c_0$, is introduced, to solve Eqs.(\ref{i2n-disc}).
The $\boldsymbol{e}_{\alpha}$ form a basis that generate an array of regularly spaced points in real space while the lattice of vectors $\boldsymbol{\xi}_{\alpha} $ does not, and so, must be locally adjusted by  $c_0$. Hereafter we restrict our study to the $N=3$ order in the EDF of Eq.(\ref{expansao0}) which renders the Navier-Stokes equation from the Chapman-Enskog analysis of the Boltzmann-BGK equation~\cite{coelho14}.
Therefore, as examples, we consider the following two lattices. The  \textit{ d1v3} ($D=1$) lattice contains only three velocities, $\boldsymbol{e}_{0}=0, \,\,\boldsymbol{e}_{\pm 1}=\pm 1$ and the associated weights are $w_0=I_0\left (1-J_2/3\right )$ and $w_{\pm 1} = I_0J_2/6$. The lattice \textit{d2v9} ($D=2$) has its geometrical velocities on the sides and diagonals of a square, namely, $\boldsymbol{e}_{\alpha}$, $\alpha=0,\cdots,8$: $\boldsymbol{e}_{0}=(0,0)$, $\boldsymbol{e}_{long}=[(1,1),(1,-1),(-1-1),(1,-1)]$ and $\boldsymbol{e}_{short}=[(1,0),(0,1),(-1,0),(0,-1)]$. The corresponding weights are $w_0=I_0\left (1-5J_2/9 \right )$, $w_l=I_0 J_2/36$ and $w_s=I_0J_2/9$, associated to the center, long and short vectors, respectively. For both lattices $c_0=\sqrt{3I_4/I_2}$. Therefore the $N=3$ EDF,
\begin{eqnarray}
&& f^{(eq)}_{\alpha}= w_{\alpha} \left \{1+\frac{I_0}{I_2}\boldsymbol{\xi}_{\alpha} \cdot \boldsymbol{u} + \frac{I_0}{2I_4}\left(\boldsymbol{\xi}_{\alpha} \cdot \boldsymbol{u}\right )^2+ \right . \nonumber \\
&& \left .\frac{I_0\left (\Delta_2^2-1 \right )}{2I_4 D}\boldsymbol{u}^2\boldsymbol{\xi}_{\alpha}^2 - \frac{I_2}{2I_4}\Delta_2^2\boldsymbol{u}^2 +\frac{1}{6}I_0\left(\boldsymbol{\xi}_{\alpha}\cdot\boldsymbol{u}\right)\cdot \right .\nonumber \\
&&  \left . \left [ 3\left( 1-J_2\right )\frac{J_4}{I_2}\left (D+2\right )\Delta_4^2-3\left(1-J_2 \right)\frac{J_4}{I_4}\Delta_4^2\boldsymbol{\xi}_{\alpha}^2-  \right . \right .\nonumber \\
&&  \left . \left . -3 \frac{J_4}{I_4}\Delta_4^2\boldsymbol{u}^2+ 3 \frac{\Delta_4^2-1}{I_6\left(D+2\right)}\boldsymbol{\xi}_{\alpha}^2\boldsymbol{u}^2
+\frac{1}{I_6}\left(\boldsymbol{\xi}_{\alpha}\cdot\boldsymbol{u}\right)^2 \right ] \right \}, \nonumber
\end{eqnarray}
becomes in the geometrical space,
\begin{eqnarray}
&& f^{(eq)}_{\alpha}= w_{\alpha} \left \{1+\frac{3}{J_2}\boldsymbol{e}_{\alpha} \cdot \boldsymbol{u}_0 + \frac{9}{2J_2}\left(\boldsymbol{e}_{\alpha} \cdot \boldsymbol{u}_0\right )^2+ \right . \nonumber \\
&& \left .\frac{9}{2}\left ( \frac{\Delta_2^2-1}{J_2 D } \right )\boldsymbol{u}_0^2\boldsymbol{e}_{\alpha}^2 - \frac{3}{2}\Delta_2^2\boldsymbol{u}_0^2 + \frac{3}{2 J_2} \left(\boldsymbol{e}_{\alpha} \cdot \boldsymbol{u}_0\right )\cdot \right . \nonumber \\
&& \left [ \left( 1-J_2 \right )J_4\left(D+2 \right)\Delta_4^2-3\left(1-J_2 \right)J_4\Delta_4^2\boldsymbol{e}_{\alpha}^2 -\right .  \\
&& \left . \left . 3J_4\Delta_4^2\boldsymbol{u}_0^2+9 \frac{J_4\left(\Delta_4^2-1\right)}{D+2}\boldsymbol{u}_0^2\boldsymbol{e}_{\alpha}^2 + 3J_4\left(\boldsymbol{e}_{\alpha} \cdot \boldsymbol{u}_0\right )^2\right ] \right \}.\nonumber
\end{eqnarray}
The lattice Boltzmann-BGK equation becomes~\cite{he97},
\begin{eqnarray}
f_\alpha(\boldsymbol{x}+ \boldsymbol{e}_\alpha,
t+1)- f_\alpha (\boldsymbol{x}, t) = - \frac{ f_\alpha(\boldsymbol{x}, t) -
f^{(eq)}_\alpha(\boldsymbol{x}, t)}{\tau}
\end{eqnarray}
where $\tau$ is the relaxation time.
In conclusion $c_0$ has become the scale velocity in replacement of $c_r$ for the two lattices considered above.
Updates are done in the density, which is equivalent to $I_0$, and in the normalized velocity $\boldsymbol{u}_0$,
\begin{eqnarray}
 I_0 = \sum_{\alpha}f_{\alpha}\boldsymbol{e}_{\alpha} \;\; \mbox{and}\;\;
 \boldsymbol{u}_0 \equiv\frac{\boldsymbol{u}}{c_0}=\frac{1}{I_0} \sum_{\alpha}f_{\alpha}\boldsymbol{e}_{\alpha}.
\end{eqnarray}
\begin{figure}[ht]
\center
\includegraphics[width=\linewidth]{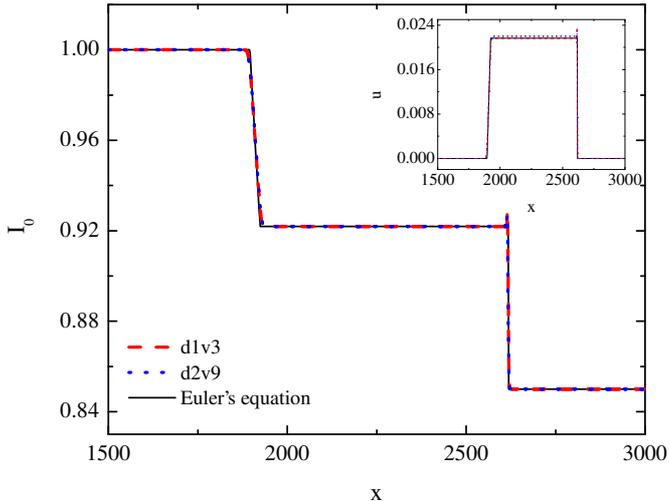}
\caption{(Color online) Simulations are done in a one-dimensional system with $L_X \times L_Y = 3001 \times 1$ nodes with periodic boundary conditions. The initial velocity is zero everywhere and the density is 1.0 for $L_X/4 < x < 3L_X/4$ and 0.85 elsewhere. Only half of the system is shown since the other one is a mirror image. The relaxation time is $\tau = 0.58$.}
\label{vrhoinset}
\end{figure}
\textbf{Numerical simulations for electrons in metals}. -- Here we develop and validate a LBM for electrons in metals based on the general semi-classical model proposed in this letter. 

The Navier-Stokes equation, 
\begin{eqnarray}
\frac{\partial}{\partial t}(I_0u^i)+ \frac{\partial}{\partial x^j}\left[I_0\left(\frac{I_2}{I_0}\delta^{ij}+u^iu^j\right)\right]-\frac{\partial \bar \sigma^{ij}}{\partial x} =0,
\end{eqnarray}
where $\bar \eta = I_2 \tau\left(1-\frac{\Delta t}{2\tau}\right)$,
is independent of $\tau$ for the one-dimensional case because the viscosity stress tensor, $\bar\sigma^{ij}= \bar \eta \left( \frac{\partial u^i}{\partial x^j}+\frac{\partial u^j}{\partial u^i}-\frac{2}{D}\delta^{ij}\frac{\partial u^l}{\partial x^l}\right)$, vanishes, for $D=1$, and so reduces to the Euler equation~\cite{coelho14}. A direct comparison between a known solution of the Euler equation in case of a stepwise initial condition, the so-called Riemann problem~\cite{toro09}, and our numerical solution of the one-dimensional Boltzmann equation is possible. We take fermions at zero temperature, that is, $\theta=0$ at eq.(\ref{sommerfeld}), which simplifies the model preserving the main physical behavior. This approximation is valid for the room temperature since it is much less than the Fermi temperature. This  renders $I_0 = (2\pi\mu)^{\frac{D}{2}}/\Gamma(\frac{D}{2}+1)$, $J_N = 1+2/(N+D)$ and $\Delta_N^2 = (D+N)/2$. Local changes in $\rho$ and $u$ are only possible due to local variations in the chemical potential. We use the two lattices, \textit{d1v3} ($w_0=7I_0/12$ and $w_{\pm 1} = 5I_0/24$) and \textit{d2v9} ($w_0=3I_0/5$, $w_l=I_0/30$ and $w_s=2I_0/15$). In this case the weights are constant in all lattice points with a remaining multiplicative density played by $I_0$ such as in the known LBM method~\cite{succi01}.
The velocity scale, $c_0=\sqrt{6\mu/(D+4)}$, is essentially the Fermi velocity. For both lattices we find independence of the numerical solution with respect to $\tau$ within a wide range ($\tau=0.57\mbox{-}0.95$, reduced units). Fig.~\ref{vrhoinset} shows the comparison between the obtained $\rho$ and $u$  with the known Riemann problem. Fermions at zero temperature obey an isothermal perfect gas law where the temperature is the Fermi temperature itself ($PV=N_0\cdot N k_B T_F$, $N_0=2/5$ in 3D). Fig.~\ref{vrhoinset} shows this comparison between the shock wave solution for an isothermal gas and the LBM simulations for the lattices
\textit{d1v3} and \textit{d2v9} after 500 steps. We find good agreement for the density and velocity curves of the \textit{d1v3} and  \textit{d2v9}. lattices. Therefore we allow for an adjustment of $N_0$ in the definition of pressure used in the solver of ref.~\cite{toro09} for the Euler equation to obtain agreement between the classical and the semi-classical gas velocities. The solver also leads to agreement in the density of the two gases.

\begin{figure}[htb]
\centering
\begin{subfigure}[h]{0.49\textwidth}
  \caption{\hfill~}
  \includegraphics[width=\columnwidth]{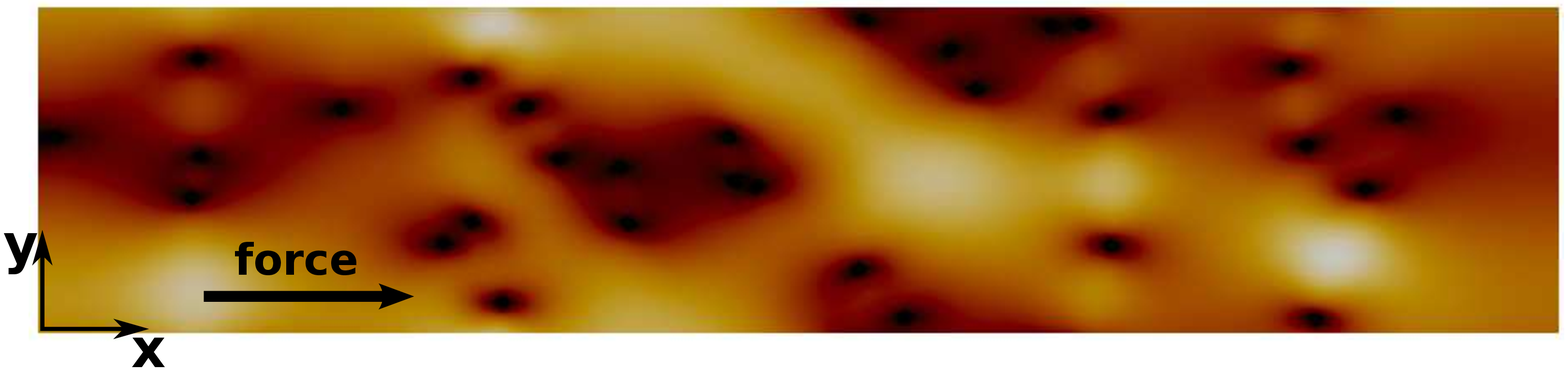}
  \label{v-field}
\end{subfigure}
\begin{subfigure}[h]{0.49\textwidth}
  \caption{\hfill~}
  \includegraphics[width=\columnwidth]{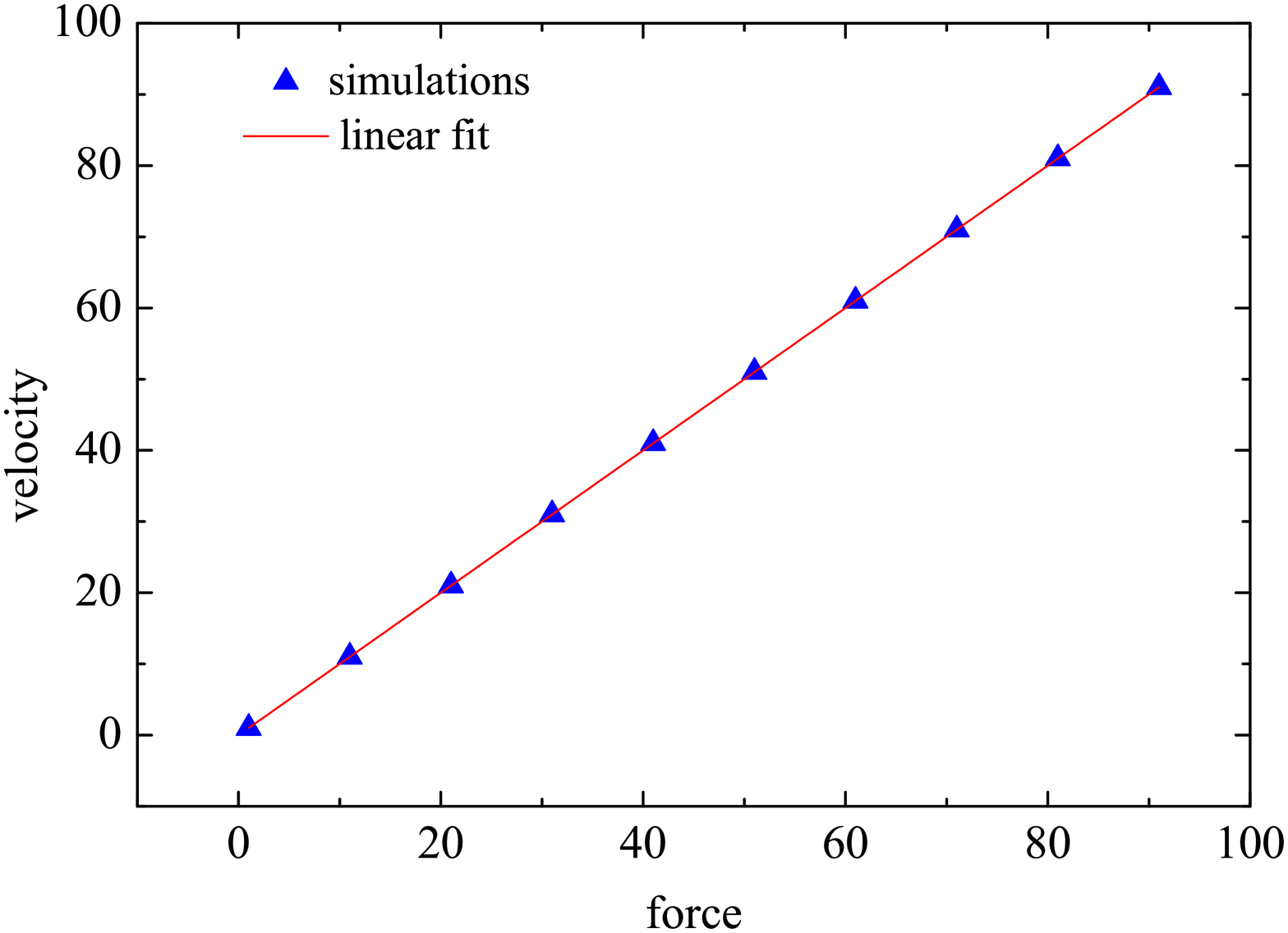}
  \label{ohm-fit}
\end{subfigure}
\caption{(Color online) Two dimensional current of electrons through obstacles in a system with dimensions of 700$\times$150 nodes. The obstacles are 30 circles with radius of 3 nodes at random positions. The relaxation time is $\tau=0.7$ in lattice units. (\ref{v-field}) Sample of the velocity norm field. The black circles are the obstacles and the gradient represents the magnitude of the velocity field, where the lighter the color, the higher the velocities. (\ref{ohm-fit}) For the same obstacles (shown in (\ref{v-field})) the force is changed and the average velocity in the direction of the force is measured. The force and velocity here are non-dimensional, i.e., both divided by the lowest value in the list. A linear fit was made, $f(x)=ax$, resulting in $a=1.00013 \pm 0.00001$.}
\label{ohm-fig}
\end{figure}

We also develop two-dimensional simulations for the electrons flowing through rigid obstacles randomly placed. Fig. \ref{v-field} shows the velocity field inside a sample of porosity $\phi = 0.992$ which we used for all simulations. The bounce-back boundary conditions were used for the obstacles and periodic boundary conditions for the borders. A constant external force acts on the fluid. In Fig. \ref{ohm-fit} we see a clear linear relations between the applied force and the average velocity in the directions of the force. This proportionality can be interpreted as the Ohms law (for electrons in metals) or as the Darcy's law (for classical fluid in a porous media)~\cite{coelho16}.

\textbf{Conclusions} -- We proposed a general approach to develop semi-classical lattice Boltzmann methods based on our discovery of generalized polynomials in D dimensions. This approach allows us to use the weight function as being the distribution function itself with the macroscopic velocity equal to zero, which improves the convergence and the accuracy. We did simulations for the Riemann problem using a LBM for electrons in metals, which is a particular case of our general model that can be useful for industry and research in condensed matter physics.    

\acknowledgments
We thank Luiz A. Hegele Jr. for helpful discussions. M. M. Doria acknowledges CNPq support from funding 23079.014992/2015-39 and R. C. V. Coelho thanks to CNPq and FAPERJ for financial support.

\bibliography{reference}

\end{document}